\documentclass[preprint,times]{emulateapj}
\usepackage{natbib}

\usepackage{graphicx}
\usepackage{epsf}
\usepackage{color} 

\begin{document}

\title{Multi-Wavelength Observations of Comet C/2011 L4 (Pan-Starrs)}
\author{Bin Yang$^{1,2}$, Jacqueline Keane$^{1}$, Karen Meech$^{1,3}$, 
Tobias Owen$^{3}$, Richard Wainscoat$^{3}$}
\affil{$1$ NASA Astrobiology Institute, University of Hawaii, Honolulu, HI 96822\\ 
$2$  European Southern Observatory, Santiago, Chile\\
$3$ Institute for Astronomy, University of Hawaii, Honolulu, HI 96822 \\} 
\email{yangbin@ifa.hawaii.edu}

\begin{abstract}

Dynamically new comet C/2011 L4 (PanSTARRS) is one of the brightest comets since the great comet C/1995 O1 (Hale-Bopp). Here, we present our multi-wavelength observations of C/2011 L4 during its in-bound passage to the inner Solar system. A strong absorption band of water ice at 2.0 $\mu$m was detected in the near infrared spectra, taken with the 8-m Gemini-North and 3-m IRTF telescopes. The companion 1.5~$\mu$m band of water ice, however, was not observed. Spectral modeling show that the absence of the 1.5~$\mu$m feature can be explained by the presence of sub-micron-sized fine ice grains. No gas lines (i.e.\ CN, HCN or CO) were observed pre-perihelion either in the optical or in the sub-millimeter. 3-$\sigma$ upper limits to the CN and CO  production rates were derived. The comet exhibited  a very strong continuum in the optical and its slope seemed to become redder as the comet approached the Sun. Our observations suggest that C/2011 L4 is an unusually dust-rich comet with a dust-to-gas mass ratio $>$ 4.  \end{abstract}

  \keywords{comets: individual (C/2011 L4)---infrared: planetary systems---Oort Cloud}

\section{Introduction}
Comet C/2011 L4 (PanSTARRS) was discovered at 7.9 AU from the Sun in June 2011 by the 1.8-m Pan-STARRS 1 survey telescope atop Haleakala. The detection was confirmed in followup observations with the Canada-France-Hawaii Telescope \citep{wainscoat:2011}. The small reciprocal of the original semi-major axis, \textit{1/a}= -8.9$\times$ 10$^{-5}$ AU$^{-1}$\citep{williams:2013}, suggesting that this comet is a recent arrival from the Oort Cloud. C/2011 L4 reached an apparent visual magnitude of \textrm{-1} at 0.301 AU in March 2013, which made it one of the brightest comets in the past two decades since comet C/1995 O1 (Hale-Bopp). The apparition of C/2011 L4 provided a rare opportunity to monitor a dynamically new comet over a significant time period and a wide range of heliocentric distances, particularly during the in-bound leg.

At the time of discovery, C/2011 L4 showed an extended appearance \citep{wainscoat:2011}, indicating the presence of a substantial coma. The turn-on distance for any appreciable water ice sublimation is around 5-6 AU from the Sun, beyond which the equilibrium surface temperatures is too low and H$_2$O-ice sublimation is not sufficient to lift even the smallest sub-micron-sized dust particles from the surface \citep{meech:2004}. It is clear that C/2011 L4 was far beyond the water sublimating zone yet it appeared active. Several possible mechanisms that could power distant comet activity have been proposed, including latent heat release from the amorphous-to-crystalline water ice transition \citep{prialnik:1992, notesco:2003, bar-Nun:2003}, sublimation of frozen super-volatiles \citep{meech:2004} and comet fragmentation \citep{boehnhardt:2004}. Among these, the exothermic amorphous-to-crystalline phase transition of water ice is considered to be an important energy source that can trigger distant cometary activity and power a cometary outburst \citep{enzian:1997}. Observationally, crystalline ice differs from its amorphous counterpart by the presence of a sharp narrow feature at 1.65~$\mu$m. This feature has been observed in large Kuiper belt objects via near infrared (NIR) spectroscopy \citep{Jewitt:2004a, trujillo:2007} and was recently detected in the coma of an outbursting Jupiter-family comet (JFC) P/2010 H2 (Vales) \citep{yang:2010}. However, crystalline water ice has never been seen in any Oort Cloud comet (OCC) \citep{davies:1997, kawakita:2004}. Nevertheless, the absence of observational evidences does not exclude the presence of crystalline ice in OCCs. This is because the available OCC spectra have limited signal-to-noise ratios (SNR) and spectral resolutions. We conducted a multi-wavelength observational campaign of C/2011 L4, from optical to sub-millimeter wavelengths. The main goals of this study were three-fold: (i) search for water ice, in particular, the crystalline feature at 1.65~$\mu$m via high SNR and medium-resolution spectroscopy in the NIR; (ii) monitor the on-set of volatile species (such as CN, HCN and CO, etc) in the optical and the radio as the comet approached the Sun, (iii) monitor the development of the dust and gas comae via optical spectroscopy.

\section{Observations and Data Reductions}
NIR spectroscopy was conducted using the GNIRS spectrograph on the 8-m Gemini North (Gemini-N) telescope and the SpeX spectrograph on the 3-m IRTF telescope. The GNIRS observations were taken with the short camera (plate scale: 0.15$^{\prime\prime}$/pix). The SXD cross-dispersed mode, 32 l/mm grating and 1.0$^{\prime\prime}$ slit, provides an averaged resolving power of R $\sim$ 500 over 0.85-2.5 $\mu$m. The SpeX observations were made using the high throughput prism mode (0.8 - 2.5 $\mu$m) and a 0.8$^{\prime\prime} \times $15$^{\prime\prime}$ slit that provides a spectral resolution of $R \sim$ 150  \citep{rayner:2003}. Two nearby G-type stars, used both as the telluric correction standard stars and solar analogs, were observed together with the comet during each run. The GNIRS data were reduced using the Image Reduction and Analysis Facility (IRAF) software and the Gemini IRAF package. The SpeX data were reduced using the reduction pipeline SpeXtool \citep{cushing:2004}. 

Optical spectra were obtained using the Gemini Multi-Object Spectrograph (GMOS) spectrograph on Gemini-N with the long-slit mode (plate scale: 0.146$^{\prime\prime}$/pix). We used the B600 grating and set the slit width to 1.5$^{\prime\prime}$, which provides a resolving power of R$\sim$ 560. At least one G-type star was observed together with the comet during each observing run. In addition, one spectrophotometric star was observed for flux calibration. The GMOS data were reduced using a combination of the Gemini IRAF package and the XIDL code (J. X. Prochaska, priv. communication).

Besides, we also obtained submillimeter spectroscopy of C/2011 L4 using the 15-m JCMT Telescope on Mauna Kea, Hawaii. The ACSIS spectrometer was used, providing a total bandwidth of 250 MHz and spectral channel spacing of 30.5 kHz. The comet was observed over three separate observing runs between August and November 2012. All of our observations were performed with the HARP receiver in position-switching mode (5$^{\prime}$ apart in azimuth). The data were reduced using a combination of Starlink and IDL software.  A record of the observations is provided in Table 1. 
 \section{Results}

\subsection{Detection of Water Ice}
The NIR observations of C/2011 L4 are presented in Figure 1. A strong absorption feature centered at 2.0~$\mu$m was consistently observed both in the SpeX and the GNIRS spectra. The profile and the center of this absorption feature is consistent with the diagnostic 2.0-$\mu$m absorption band of water ice. However, water ice, if present, usually exhibits two absorption bands simultaneously, i.e.\ the 2.0~$\mu$m band and an accompanying band at 1.5~$\mu$m. The spectra obtained in March and June appear featureless between 1.4 and 1.7~$\mu$m. The July spectrum, however, shows two small absorption features at 1.28 and 1.50~$\mu$m, respectively. The latter feature could be associated with water ice but it is significantly narrower than the 1.5 $\mu$m ice band. Comparing the comet spectrum with the standard star spectra taken in July, we conclude that these two minor features are more likely due to incomplete cancellations of telluric absorptions. The apparent absence of the 1.5~$\mu$m feature in contrast to the presence of the strong 2.0~$\mu$m band is not unique to C/2011 L4. For instance, the outbursting comet 17P/Holmes exhibited strong absorption bands at 2.0 and 3.0~$\mu$m without showing any sign of the 1.5~$\mu$m band \citep{yang:2009}. Attempting to understand the problem of the missing 1.5$\mu$m feature, \cite{yang:2009} investigated the effects of impurity and particle size on the synthesized ice spectra. They concluded that neither of the investigated properties could explain the asymmetry between the 1.5 and 2.0~$\mu$m bands. However, in the study of 17P, the adopted Hapke theory has significant limitations and cannot account for particles smaller than the wavelength being studied \citep{hapke:1993}. In this paper, we apply Mie theory to synthesize water ice spectra and the optical constants from \cite{Mastrapa:2009}. The main reason for choosing Mie over Hapke is that Mie theory is suitable for investigating small particles that have Rayleigh scattering parameter $X$=2$\pi a_g / \lambda$ much less than one, where $a_g$ is particle radius and $\lambda$ is the wavelength. 

Using the new Mie model, we found that the ratio between the depth of the 1.5 and 2.0~$\mu$m bands is sensitive to grain size. Using sub-micron fine grains, our best-fit areal mixing models can successfully reproduce the observed spectral features, i.e.\ a strong 2.0 $\mu$m band and a much weaker 1.5 $\mu$m band. All the comet spectra can be explained by roughly 30\% fine-grained water ice ($a_g\sim$0.2$\mu$m) mixed with spectrally featureless materials (such as amorphous carbon). We note that the strength of the 2.0~$\mu$m band remained consistent from March to July, indicating that the dust-to-ice ratio of the coma did not change significantly over heliocentric distances from 5.24 to 3.68 AU. The persistence of the ice features suggests that the lifetime of the ice particles must be longer or comparable to the slit-crossing time. For the July run, the extraction aperture radius is 2.5$^{\prime\prime}$ (or $\sim$6700 km projected in the sky). Assuming the expansion speed of the icy grains is represented by the empirical relation: $v_{dust}$=0.535$\cdot r_h^{-0.6}$ km/s \citep{bobrovnikoff:1954, whipple:1978}, the slit-crossing time is at least 7.6 hours. As shown in \cite{lien:1990} and \cite{beer:2006}, the lifetime of a sub-micron dirty ice grain at 3.68~AU is less than 2 hours. If dirty icy particles were to survive the slit-crossing, their outflow velocity would have to exceed 1~km/s, which is non-physical for the comet at $r_h > $ 3.5 AU. Given the estimates above, the ice particles detected in comet C/2011 L4 it is likely to be pure. 

\subsection{Optical Spectroscopy}
Optical spectroscopy was obtained using Gemini-N in queue mode from April to July 2012. Specifically, we searched for the CN 3880\AA\ emission band as a probe to monitor the sublimation of volatile species. After subtracting the scaled solar analog continua, the residual spectra appear featureless with no apparent cometary emission lines detected (see Figure 3a). Upper limits to the CN production rate were quantified using the standard errors in the two wavelength regions, 3900-3950\AA\ and 3950-4010\AA , of the residual spectra. Details of our method are described in \cite{hsieh:2012}. 3-$\sigma$ limits to the CN band flux and the CN production rate are listed in Table 2. The derived upper limits (5.0-6.0$\times$10 $^{27}$ mole/s) to the water production rate of C/2011 L4, are consistent with the H$_2$O production rate of 5.1$\times10^{27}$ mol/s  in September 2012, reported by \cite{biver:2012}. 

Since no gas emission was detected, the continuum must be solely controlled by the dust coma. To assess the strength of the dust continuum, we calculate $Af\rho$ \citep{ahearn:1984} and list our results in Table 2. The $Af\rho$ of C/2011 L4 between 4900 and 5500\AA\ exhibits small but steady increases from April to July, which are consistent with an independent photometric study of this comet \citep{Ivanova:2014}, obtained during a similar period of time. In terms of $Af\rho$,  C/2011 L4 dwarfs the majority of observed OCCs \citep{storrs:1992, langland-shula:2011}. However, it is an order of magnitude lower than that of C/1995 O1 \citep{weiler:2003}. We further estimated the dust mass loss rate, using the formulations in \cite{weiler:2003} and \cite{agarwal:2007}. Given that the gas production rate was low and our NIR observations suggest the presence of sub-micron ice particles, we assumed a differential size distribution of f(a) $\sim$ a$^{-4}$, which gives an effective mean dust size: $\bar{a}=\frac{\int \pi a^3 f(a)}{\int \pi a^2 f(a) } < $ 1 $\mu$m. For dust velocity, we adopted an outflow velocity relationship of $v_{dust}(a)$=0.5*$\sqrt{\frac{C \cdot Q_s}{\rho_d \cdot a}}$ km/s from \cite{lisse:1998}, where C=0.595$\times$ 10$^{-3}$ kg m$^{-2}$, the scattering efficiency $Q_s \sim$ 1 and the dust bulk density $\rho_d$=10$^{3}$ kg m$^{-3}$. Integrating over a size interval from 0.1$\mu m$ to 1$cm$ and assuming a geometric albedo of 0.04 for dust particles \citep{singh:1992}, we compute the dust mass-loss-rate of C/2011 L4 to be about 700 kg/s. Using the upper limit to the CN production, we estimated that the upper limit to the gas mass-loss-rate is $\sim$ 200 kg/s. Shown in Table 2, the dust-to-gas mass ratio ($\chi$) of C/2011 L4 is $>$ 4 in July 2012. In comparison, C/1995 O1 holds the highest $\chi$ that is $\gtrsim$ 5 \citep{jewitt:1999} while most comets have $\chi$ less than 2 \citep{singh:1992, sanzovo:1996}. Our results indicate that C/2011 L4 is an unusually dust-rich comet. We note that the dust size distribution and the dust outflow velocity are the most uncertain factors in calculating the dust production rate. If we assume f(a) $\sim$ a$^{-3.5}$, the corresponding dust-to-gas ratio would be about 100, which is highly unlikely. Detailed models that fully explore the parameter space for these two factors are beyond the scope of this paper. 

\subsection{Submillimeter Observation}
In the radio, CO and HCN are the most easily detected gaseous species for distant comets (r $>$  3 AU) \citep{senay:1994, biver:2002}. We carried out three observing runs in the sub-millimeter to search for CO and HCN. No gas emission was detected from August to November. To quantify the upper limits to the production rates of CO and HCN, we calculated the standard errors in two background regions and integrated line intensities ($f_{CO}$ and $f_{HCN}$) within a 1.2 km s$^{-1}$ band. Even at the smallest $r$ (2.3 AU), where the CO lifetime is the shortest, the expected lifetime against photo-destruction for a CO molecule is $\tau_{CO}$ = 3.6 $\times 10^{6}\cdot r_h^2$=1.9 $\times 10^{7}$ sec (about 1 year). Similarly, $\tau_{HCN}$ = 4.8 $\times 10^{5}$ sec (about 4 days). Therefore, we ignored photo-destruction effect for both CO and HCN. We assumed gas molecules escaping from the surface at a constant velocity and adopted an average gas expansion velocity of 0.8 $\cdot$r$ ^{-0.5}$ km s$^{-1}$ (Biver et al., 1999, 2000). 3-$\sigma$ upper limits to the production rates were computed for two kinetic temperatures, i.e T= 10~K (higher values) and T = 50~K (lower values). For August, $f_{CO}$= 0.015 [K km/s], Q$_{CO} < $ 1.5-4.6$\times 10^{27}$, $f_{HCN}$= 0.032 [K km/s], Q$_{HCN} < $0.5-3.1$\times10^{25}$; for October, $f_{CO}$= 0.024 [K km/s], Q$_{CO} < $2.4-7.4$\times10^{27}$, $f_{HCN}$= 0.035 [K km/s], Q$_{HCN} < $0.5-3.3$\times10^{25}$;  and for November, $f_{CO}$= 0.093 [K km/s], Q$_{CO} < $ 0.9-2.8$\times10^{28}$, $f_{HCN}$= 0.037 [K km/s], Q$_{HCN} < $0.6-3.6$\times10^{25}$. The much higher upper limit on Q$_{CO}$ derived for the November run is due to poor weather conditions. 

\section{Discussion}
\subsection{Water Ice Features}
The critical temperature for amorphous-to-crystalline ice transition is T$_c$ $\sim$ 140 K, at which the timescale of crystallization is about 1 hour. The highest possible surface temperature for C/2011L4 should be that at its subsolar point, where T$_{ss}$(r$_h$)=T$_0$/$\sqrt{r_h}$. If we assume C/2011 L4 is a slow rotator, then T$_0$=392 K \citep{jewitt:2009}. At r$_h$=5.2 AU, C/2011 L4 can reach T$_{ss}\sim$170K by solar heating alone. Although this is an extreme case, it shows that it is possible that at least part of the surface or near-surface ice could have experienced crystallization within our observation window.

 As shown in Figure 2, sub-micron-sized ice grains show a very weak or non-existent absorbing feature at 1.65~$\mu m$ regardless of the physical structure of individual grains. Thus, the absence of the 1.65~$\mu m$ band does not exclude the presence of very fine crystalline ice grains. When focusing on the 2.0~$\mu$m band, our $\chi^2$-fit show that amorphous ice fits the comet spectra better than crystalline ice for the March and June data, while the crystalline model fits the July data marginally better. However, the small difference between these two ice models is within the noise level of our data. Therefore, we cannot constrain the crystallinity of the icy grains of C/2011 L4, mainly due to their small sizes. 

Although the observed asymmetry in absorption strength between the bands at 1.5~$\mu m$ and 2.0~$\mu m $ can be successfully explained by sub-micron pure water ice grains, there are other explanations for the observed features. For example, the 2.0 $\mu m$ absorption feature may not be solely generated by pure water ice grains but in addition be due to the superposition another absorption band of organic materials or hydrated minerals. We explored this possibility by searching for matching materials that exhibit a prominent absorption feature at or near 2.0~$\mu$m. We investigated several commonly used spectral libraries, such as RELAB \citep{hiroi:1996}, USGS \citep{clark:2003} and ASTER \citep{baldridge:2009}, which include examined materials from terrestrial minerals to meteorite samples. We find no other material with a NIR signature that matches the shape and position of the 2.0 $\mu m$ band of C/2011 L4.

\subsection{Dust Coma}
Shown in Fig 3b, the slope of the dust continuum became steeper as the comet approached the Sun. Similar color-$r_h$ trends have been reported in other comets \citep{hartmann:1984}. However, later studies \citep{jewitt:1988} did not confirm these reports, suggesting instead that the reported trend may have been due to the contamination of cometary emissions. Given that no cometary emissions were observed in our optical data, the observed color trend is not caused by gas contamination. Possible explanations for this correlation, if real, include: (a) increase in the effective size of coma grains, (b) decrease in the ice-to-dust ratio in coma grains and (c) changes in phase angle as the comet approaches the Sun. Our NIR observations suggest that the ice-to-dust ratio remained more or less the same (within uncertainties) and the phase angle of C/2011 L4 did not change much from April to July. Therefore both (b) and (c) scenarios are unlikely and we are left with (a) as the only plausible mechanism: the observed color-$r_h$ trend is observed because larger dust particles being released into the coma as the comet moved closer to the Sun.

Given that no gas was detected, what could be the driving force for the cometary activity? One possibility is that C/2011 L4 underwent a strong outburst which released most of the observed dust and ice grains into the coma. At the time of our observations, the major gas production had ceased and the nucleus was surrounded by a remnant dust cloud. However, the change in the slope of the dust continuum argues for an on-going mass loss process. C/2011 L4 is a dynamically new comets, so reactive species could have been synthesized via long term exposure to cosmic rays or galactic UV bombardment while the comet was in the Oort Cloud \citep{hughes:1990}. These radical species could explode when exposed to the solar wind and that could in turn trigger several small outbursts \citep{niedner:1980}. Alternatively, the distant activity of C/2011 L4 could be driven by highly volatile ices, such as CO$_2$. Recently, several comets, such as comet 103P/Hartley 2 and comet ISON, haven observed to show strong CO$_2$ signatures \citep{ahearn:2011, lisse:2013}. Unfortunately, CO$_2$ can not be observed from the ground and no space observations are available for C/2011 L4. To verify whether CO$_2$ could be driving the cometary activity, evolutionary models \citep{meech:2013} which attempt to match the brightness measurements of C/2011 L4 are needed.

In summary, our multiwavelength observations yield the following findings: 1) very fine water ice grains in the coma of C/2011 L4; 2) no gas emission was detected; 3) C/2011 L4 is unusually dust-rich; 4) the dust coma seemed to becoming redder with decreasing r$_h$.

\section{Acknowledgment}
We thank Jason X. Prochaska and Kate Rubin for helping us with the GMOS data reduction and Zahed Wahhaj, David Jewitt for valuable discussions and constructive suggestions. We thank the JCMT staff for their assistance. B.Y. was supported by the NASA Astrobiology Institute under Cooperative Agreement No.\ NNA08DA77A issued through the Office of Space Science. Work was supported in part by NSF grant AST-1010059.


\newpage
\begin{deluxetable}{llclllc}\tablewidth{6.5in}
\tabletypesize{\scriptsize}
\tablecaption{ Observational Parameters for C/2011 L4 (Panstarrs)
  \label{obstable}}
\tablecolumns{6} \tablehead{  \colhead{UT Date }  & \colhead{$r_h$ \tablenotemark{a}}  & \colhead{$\Delta$ \tablenotemark{b}} &   \colhead{ $\alpha$ \tablenotemark{c}}  &  \colhead{Tel.} &  \colhead{Inst.}& \colhead{$\lambda$ \tablenotemark{d}}  \\
\colhead{}&\colhead{(AU)}&\colhead{(AU)}&\colhead{deg}&\colhead{}&\colhead{}&\colhead{$\mu$m}}
\startdata
2012 Mar.\ 18 &  5.248 &  4.956 &10.71 & IRTF & SpeX & 0.80 - 2.5 \\
2012 Apr.\ 22 &  4.882 & 4.081 & 7.80 & Gemini & GMOS & 0.38 - 0.6 \\
2012 May 15 &  4.635 & 3.650 & 3.21 & Gemini & GMOS  & 0.38 - 0.6\\
2012 Jun.\ 06 & 4.392 & 3.400 & 3.13 & Gemini & GNIRS & 0.85 - 2.5  \\
2012 July.\ 12 & 3.980 & 3.328 & 12.30 & Gemini & GMOS & 0.38 - 0.6  \\
2012 July.\ 19 & 3.898 & 3.350 & 13.61 & IRTF & SpeX& 0.80 - 2.5 \\
2012 Aug.\ 25-27 & 3.437 & 3.526 & 16.63  & JCMT & HARP& 800-922  \\
2012 Oct.\ 20-22 & 2.713& 3.520 & 10.49  & JCMT & HARP& 800-922  \\
2012 Nov.\ 19-21 & 2.310 & 3.256 & 5.62 & JCMT & HARP& 800-922  \\
\enddata
\tablenotetext{a}{Heliocentric distance}
\tablenotetext{b}{Geocentric distance.}
\tablenotetext{c}{Solar phase angle (Sun-object-Earth).}
\tablenotetext{d}{Wavelength coverage.}
\end{deluxetable}

\begin{deluxetable}{lccccccl}\tablewidth{6.0in}
\tabletypesize{\scriptsize}
\tablecaption{ $Af\rho$ and dust production rates for C/2011 L4 (Panstarrs)
  \label{obstable}}
\tablecolumns{8} \tablehead{ \colhead{UT Date }& \colhead{$f_{CN}$\tablenotemark{a}}  & \colhead{Q$_{CN}$\tablenotemark{b}} &\colhead{Q$_{H_2O}$\tablenotemark{c}} &
\colhead{Af$\rho$}  & \colhead{$\dot{M}_d$\tablenotemark{d}}  &  \colhead{$\dot{M}_g$\tablenotemark{e}} &  \colhead{$\chi$\tablenotemark{f}}\\
\colhead{}&\colhead{($erg/cm^{2}/s/\AA$ )}&\colhead{(mol/s)}&\colhead{(mol/s)}&\colhead{(m)}&\colhead{(kg/s)}&\colhead{(kg/s)}&\colhead{}}
\startdata
2012 Apr.\ 22 &  1.51$\times 10^{-15} $  & $<$ 1.68$\times 10^{25}$ & $< $ 5.9$\times 10^{27}$ & 110.2 $\pm$ 8.0& 670 & $< $ 200  & $> $ 3.4  \\
2012 May 15 &  1.92$\times 10^{-15} $ &  $< $ 1.78$\times 10^{25}$ & $< $ 6.2$\times 10^{27}$ & 116.9 $\pm$ 8.0& 650 & $< $ 212 & $> $ 3.1 \\
2012 July.\ 12 & 1.90$\times 10^{-15} $  & $< $ 1.41$\times 10^{25}$ & $< $ 5.0$\times 10^{27}$ & 128.9 $\pm$ 9.0 & 730 & $< $ 168  & $> $ 4.3  \\
\enddata
\tablenotetext{a}{3$\sigma$ upper limit of the CN band flux}
\tablenotetext{b}{CN production rate, upper limit}
\tablenotetext{c}{H$_2$O production rate, upper limit, assuming Q$_{CN}$/Q$_{H_2O} \sim$ 350 \citep{ahearn:1995}}
\tablenotetext{d}{Dust mass loss rate, assuming a size distribution of f(a) $\sim$ a$^{-4}$}
\tablenotetext{e}{Gas mass loss rate, assuming $\dot{M}_g$=3.4$\times$ 10$^{-26}$ Q$_{H_2O}$ \citep{sanzovo:1996}.}
\tablenotetext{f}{Dust-to-gas ratio.}

\end{deluxetable}

\begin{figure}[h]
\begin{center}
\vspace{0.5 cm}
\hspace{-1 cm}\includegraphics[width=5in,angle=0]{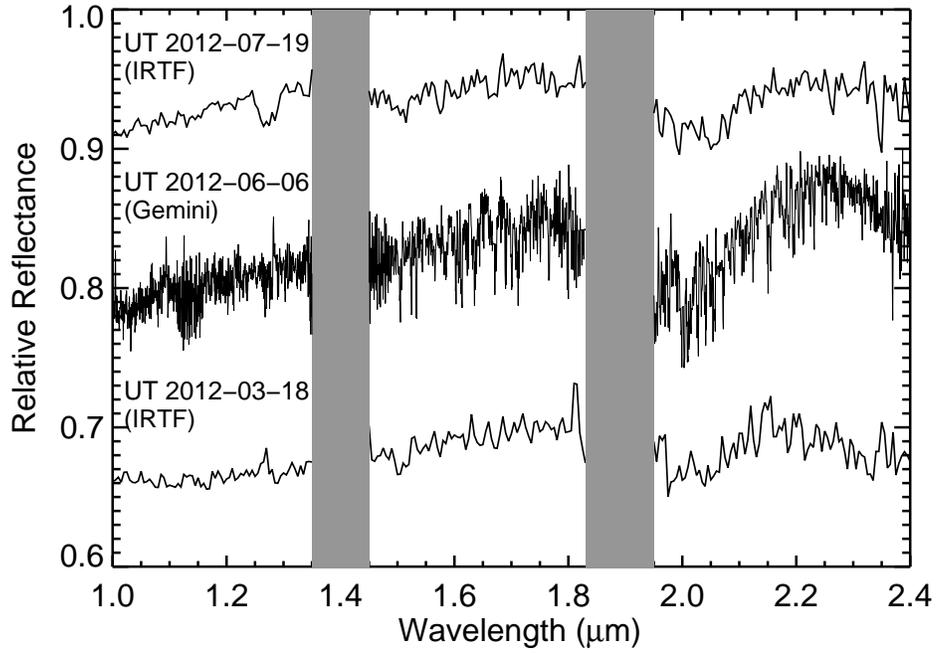}
\caption{NIR spectra of C/2011 L4, taken with the Gemini-N and IRTF telescopes. All the spectra consistently show reddish continuum and strong absorption feature centered at 2.0 $\mu$m. The Gemini spectrum appears noisier than the IRTF spectra because the former has a higher spectral resolution (R$_{G}\sim$ 560 ) than the latter (R$_{I}\sim$ 150). Spectral regions that are contaminated by atmospheric absorptions are blocked by grey bars. In the March data (bottom panel), the emission-like feature at 1.82 and 2.15$\mu$m in the March data are artifacts due to telluric absorption.}
\end{center}
\label{all_fig1}
\end{figure}

\begin{figure}[h]
\begin{center}
\vspace{-1.5 cm}
\includegraphics[width=4.in,angle=0]{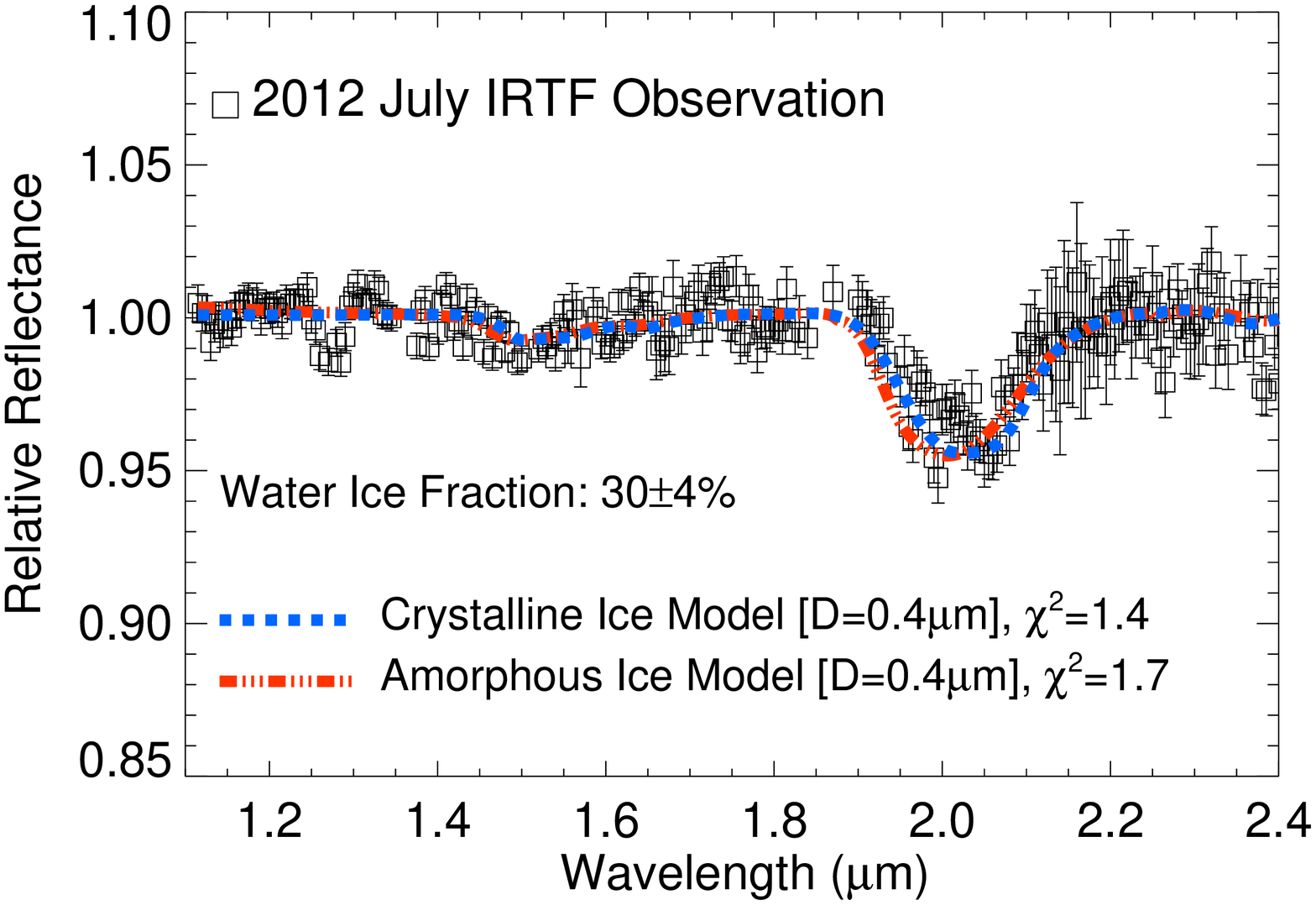}\\
\includegraphics[width=4.in,angle=0]{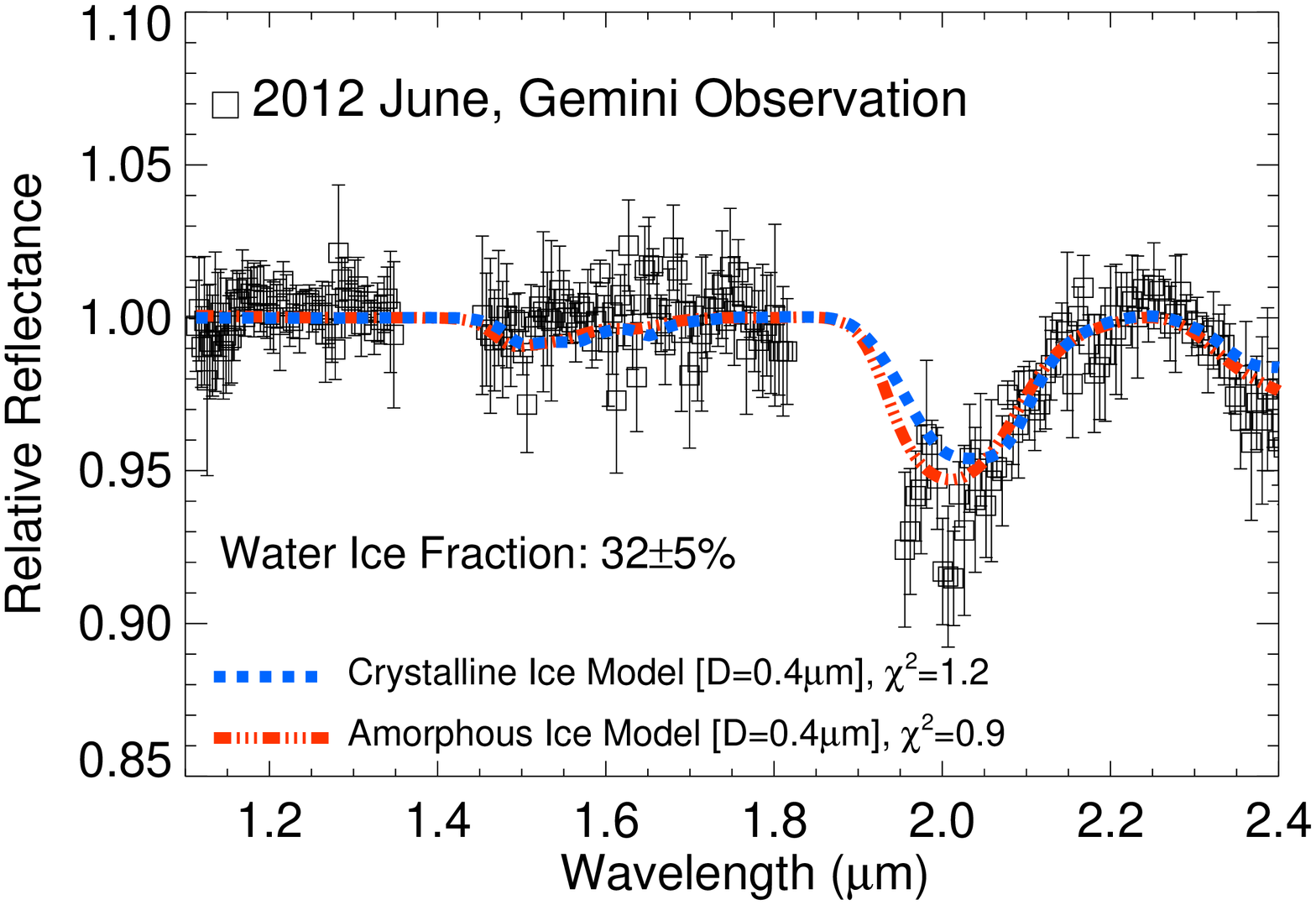}\\
\includegraphics[width=4.in,angle=0]{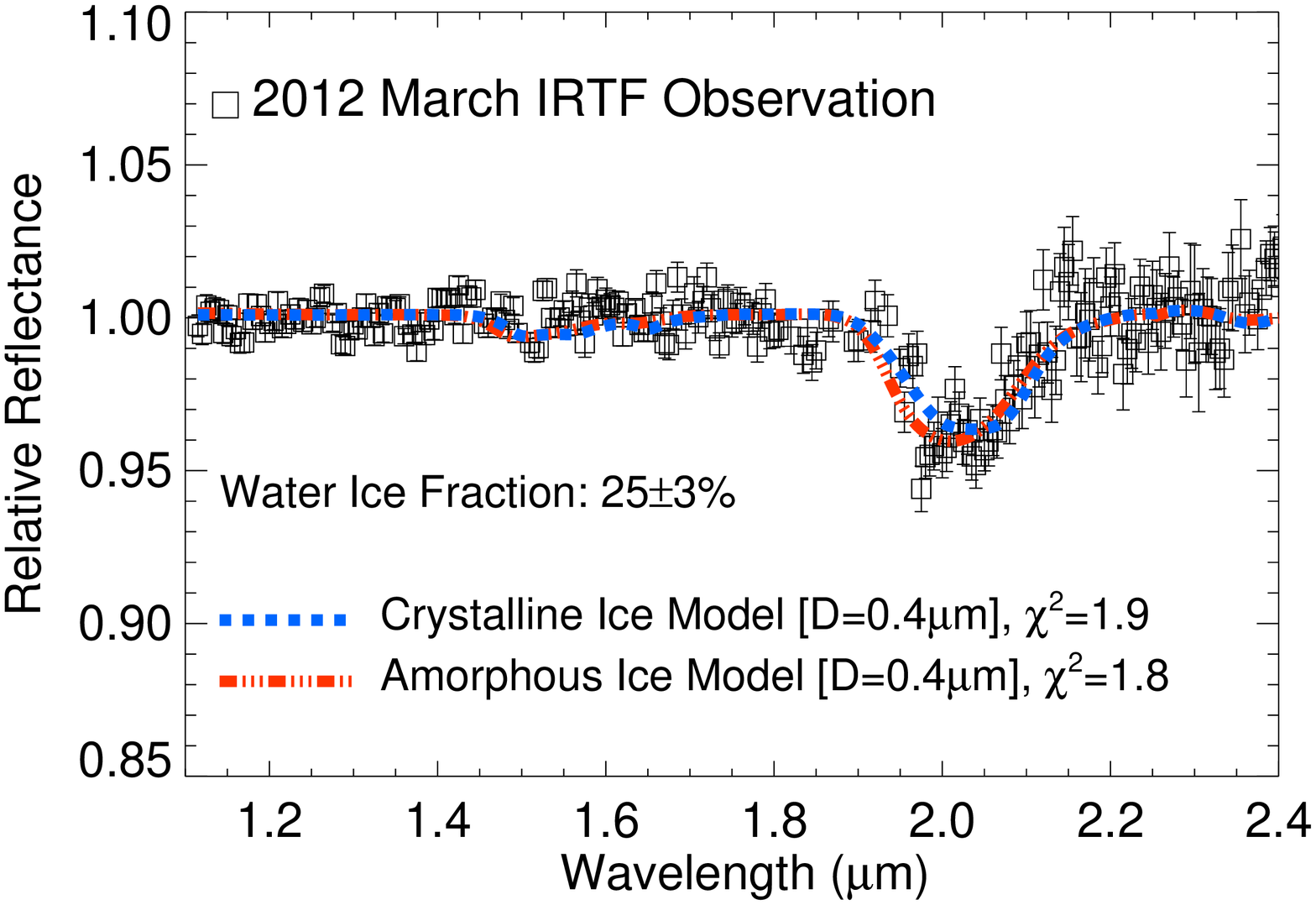}\\

\caption{The open squares are the continuum-removed NIR spectra of C/2011 L4. The red and the blue dashed lines are simple spectral models using pure amorphous water ice and crystalline ice mixed with spectrally neutral material. The model is fitted by minimizing chi-square with an iterative procedure. The chi-square values are calculated from 1.95 to 2.2$\mu$m, focusing on the fit to the 2.0$\mu$m band. The discrepancy seen at 2.0$\mu$m is due to imperfect removal of a CO$_2$ absorption band due to Earth's atmosphere. }
\label{ice_model}
\end{center}
\end{figure}

\begin{figure}[h]
\vspace{-1.5 cm}
\includegraphics[width=3.7in,angle=0]{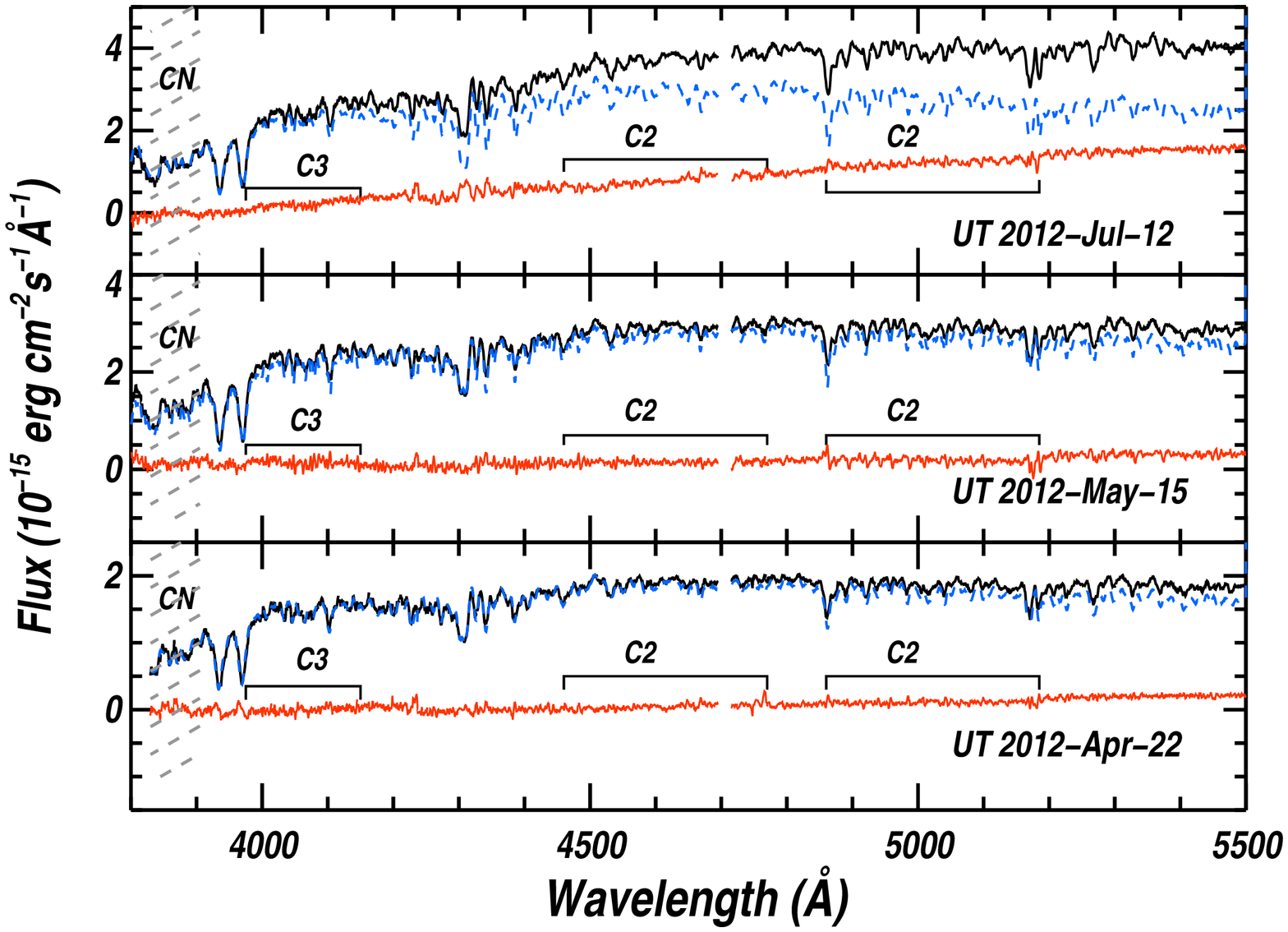}\hspace{-0.5 cm}\includegraphics[width=3.7in,angle=0]{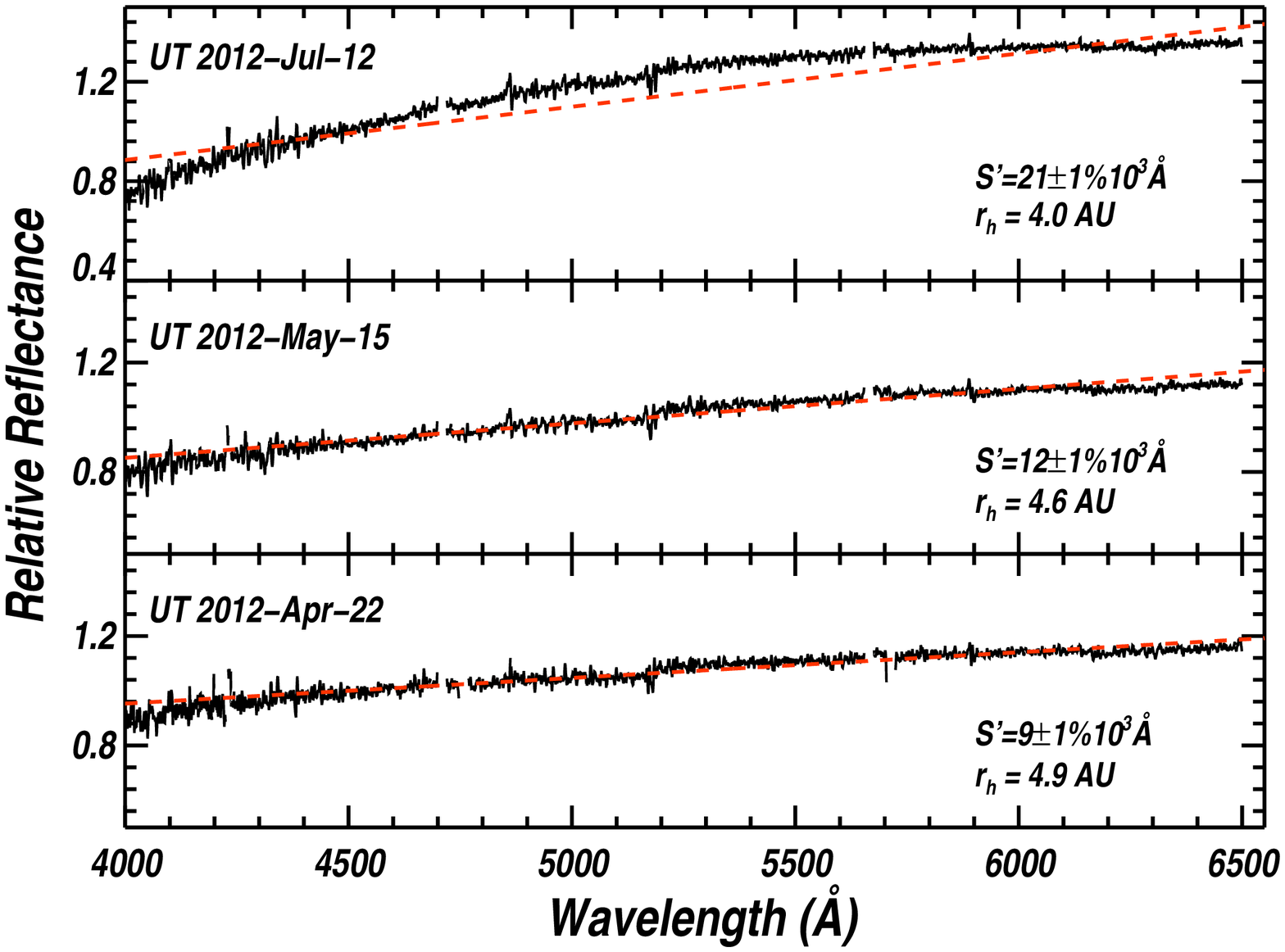}
\caption{(a) Flux calibrated optical spectra of C/2011 L4 are shown in black. Blue lines are the scaled solar analog spectra and red lines are residuals after removing the solar analog continuum. The expected CN, C$_3$ and C$_2$ bands are marked, the bandpasses are taken from \citep{cochran:1992}. Small emission-like features in the residual spectra are artifacts due to imperfect removal of the telluric absorptions. (b) Reflectance spectra of C/2011 L4 (black) and linear regression fits (red). The spectral slope seems to be correlated with the heliocentric distance, i.e.\ the comet spectrum becomes redder as the heliocentric distance decreases. Two gaps at 4700\AA\ and 5700\AA\ are caused by $\sim$ 2.8$^{\prime\prime}$ gaps between the detector chips of GMOS.}
\label{gmos_spec}
\end{figure}

\end{document}